\documentclass[12pt]{elsarticle}
\usepackage{amssymb}
\usepackage{esint}

\journal{Flow measurement and instrumentation}

\begin{document}

\begin{frontmatter}



\title{Visualization of the global flow structure in a modified 
Rayleigh-B\'enard setup using contactless inductive flow tomography}


%\author{Thomas Wondrak\corref{cor1}\fnref{label2}}

\author[HZDR]{Thomas Wondrak}

\ead{t.wondrak@hzdr.de}

\author[HZDRI]{Josef Pal}

\ead{j.pal@hzdr.de}

\author[HZDR]{Frank Stefani}

\ead{f.stefani@hzdr.de}

\author[HZDR]{Vladimir Galindo}

\ead{v.galindo@hzdr.de}

\author[HZDR]{Sven Eckert}

\ead{s.eckert@hzdr.de}

\address[HZDR]{Helmholtz-Zentrum Dresden-Rossendorf, Bautzner Landstr. 400, 01328 Dresden, Germany}

\address[HZDRI]{HZDR Innovation GmbH, Bautzner Landstr. 400, 01328 Dresden, Germany}

%\author{}

%\address{}

\begin{abstract}
Rayleigh-B\'enard convection is not only a classical problem 
in fluid dynamics but plays also an important role in many 
metallurgical and crystal growth applications. The measurement of 
the flow field and of the dynamics of the emerging large-scale 
circulation in liquid metals is a challenging task due to 
the opaqueness and the high temperature of the melts. Contactless 
inductive flow tomography is a technique to visualize the mean 
three-dimensional flow structure in liquid metals by measuring 
the flow induced magnetic field perturbations under the 
influence of one or several applied magnetic fields. In this 
paper, we present first measurements of the flow induced 
magnetic field in a Rayleigh-B\'enard 
setup, which are also used to investigate 
the dynamics of the large-scale 
circulation. Additionally, we investigate 
numerically the quality of the reconstruction of the 
three-dimensional flow field for 
different sensor configurations.

\end{abstract}


\end{frontmatter}


\section{Introduction}
\label{intro}

Buoyancy driven flows of  fluids heated from below and
cooled from above play a key role in geo- and 
astrophysics, in the natural ventilation of buildings, 
and in many metallurgical applications. 
It is thus no surprise that Rayleigh-B\'enard (RB)
convection has served as one of the central paradigms 
for stability theory, pattern formation,
and scaling behavior \cite{Kadanoff2001}.

Besides asking how global thermohydraulic
features such as the Reynolds and the Nusselt number 
depend on the Rayleigh 
and the Prandtl number \cite{Ahlers2009}, 
special focus of RB studies has been laid on
the phenomenon of large-scale convection (LSC)
\cite{Krishnamurti1981,Sano1989,Takeshita1996,Cioni1997,XiLamXia2004,BrownAhlers2006,Resagk2006,Xi2008}.
LSC appears at sufficiently high 
values of the Rayleigh number, when thermal plumes
erupt from the boundary layers and self-organize 
into a flywheel structure \cite{Kadanoff2001}.
The dynamics of this global wind has turned out 
surprisingly rich, comprising torsional modes \cite{Funfschilling2004} 
and sloshing modes \cite{XiZhou2009,BrownAhlers2009}, 
as well as global reorientations by azimuthal 
rotations and (rare) cessations \cite{BrownAhlers2006,Xi2008}. 
A number of
experiments dedicated to these LSC problems were carried
out with water \cite{Krishnamurti1981,BrownAhlers2006,Xi2008}, 
silicon oil \cite{Krishnamurti1981}, helium-gas \cite{Sano1989}, 
air \cite{Resagk2006}, 
and liquid mercury \cite{Takeshita1996,Cioni1997}. 

RB experiments with liquid metals are not only
fundamentally important to explore the low Prandtl 
number regime, but are also relevant 
for a variety of  
metallurgical and crystal growth  applications. 
For instance, in the Czochralski (Cz) crystal growth
of mono-crystalline silicon large 
temperature gradients are inherently present at 
the crystallization edge which lead to strong temperature 
fluctuations \cite{Muiznieks2007,Cramer2010}. The control 
(and optimization) of these fluctuations still remains a 
challenge. The flow structure, which plays a key role 
for the quality 
of the final crystal since 
it controls the temperature gradient 
and mass transport, is 
governed by the details of heating and the differential 
rotation of crucible and crystal. For stabilizing the 
flow, external magnetic fields are frequently used 
\cite{Muiznieks2007}. 

While dedicated RB experiments with low-melting liquid metals
such as mercury and GaInSn can easily be equipped with
direct contact sensors such as 
thermistors \cite{Cioni1997} or ultrasonic transducers
\cite{Tasaka2015}, industrial 
high temperature applications 
(e.g. with liquid silicon) demand for contactless 
flow measurement techniques.

A step in this direction is the contactless inductive 
flow tomography (CIFT) \cite{Stefani2000,Stefani2004,Ratajczak2014}, 
which is able to provide a picture of the 
three-dimensional flow by 
applying primary magnetic fields to the melt and by 
measuring the flow induced magnetic field perturbation 
outside the fluid volume. From these field perturbations 
CIFT infers the flow field by solving a linear inverse 
problem. Appropriate regularization techniques, 
like Tikhonov regularization in combination with the 
L-curve technique \cite{Hansen1992}, 
are necessary in order to mitigate 
the intrinsic 
non-uniqueness of the inverse problem. 
In a first demonstration 
experiment \cite{Stefani2004}, the propeller 
stirred three-dimensional flow of the eutectic 
alloy GaInSn in a cylindrical vessel was reconstructed. 
The CIFT-inferred mean velocity turned out to be in 
good agreement with comparative ultrasonic Doppler flow 
measurements.
Later, CIFT was successfully applied to visualize the
(essentially two-dimensional) flow field in the mold
of a laboratory model of continuous slab casting 
\cite{Wondrak2010,Wondrak2011,Wondrak2012}.

As for Czochralski crystal growth,  
any real application of CIFT to determine the flow 
in a Cz-puller would face a number of problems. Due 
to the high temperature of about 1500$^{\circ}$\,C in 
the oven, the magnetic field sensors can only be placed 
outside the device, which necessarily 
results in a large distance 
between the sensors and the melt. To 
make matters worse, the typical meridional flow velocity 
is only in the order of a few cm/s, i.e. 
one order of magnitude smaller than in the case of 
continuous casting. Both 
facts would decrease significantly the ratio between the 
measurable induced field and the applied magnetic 
field, which is a crucial criterion for the technical 
applicability of CIFT. 

\begin{figure}[!htbp]
	\centering
	\includegraphics[width=0.99\textwidth]{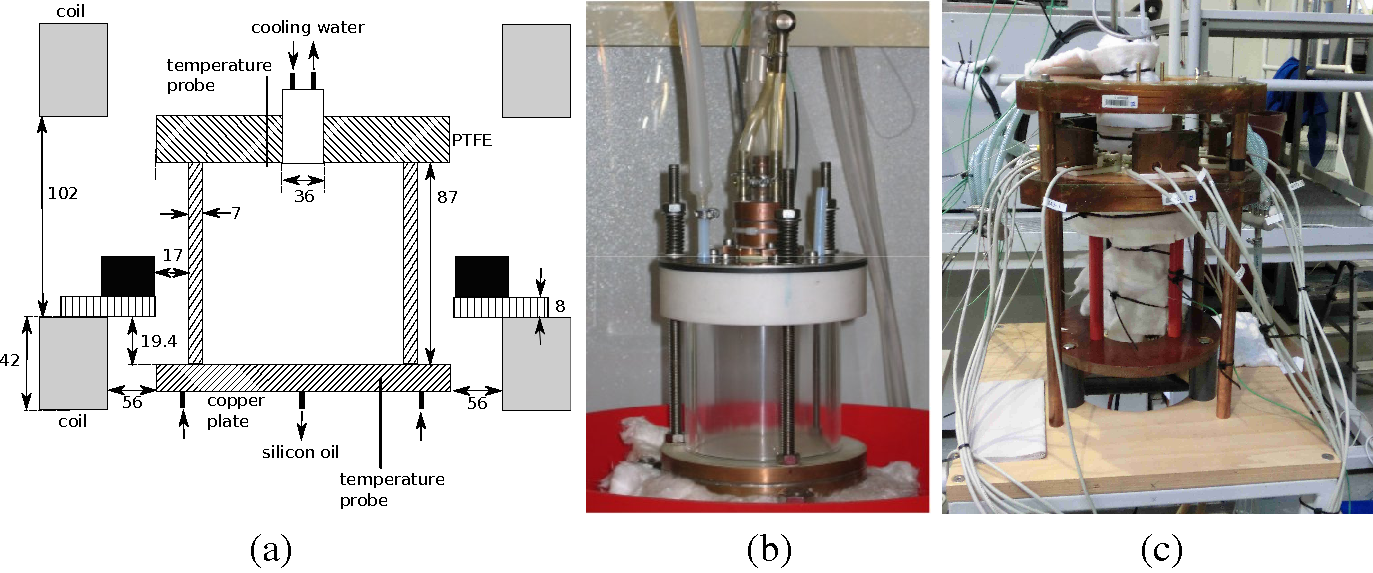}
	\caption{
	Schematic sketch (a), photography of the cell (b), 
	and photography of the complete set-up (c) 
	of the RB experiment.}
	\label{Fig:setup}
\end{figure}

Motivated by this challenge, this paper is intended 
to examine the viability of CIFT for measuring liquid metal 
velocities in the order of 1 cm/s as they are 
typical for liquid metal RB convection in general, and for 
Czochralski crystal growth in particular.
For this purpose, an existing set-up of a modified RB 
convection cell  \cite{Cramer2010} 
has been equipped with one 
pair of excitation coils, generating a basically
vertical magnetic field in the fluid, and 20 
magnetic field sensors which are situated along 
the azimuth, approximately at mid-height of the 
convection cell (Fig.~\ref{Fig:setup}).

We will demonstrate that typical dynamical features of 
the LSC, like 
azimuthal rotations and even cessations, as well as
the typical frequencies
of the torsional/sloshing mode  
\cite{BrownAhlers2009} 
are well detectable. 
While our main goal is to proof the principle applicability of 
CIFT for RB flows, a full three-dimensional flow 
reconstruction would require more sensors 
at different 
heights, and, perhaps, a second magnetic 
field to be applied in horizontal direction. 
Therefore, we will 
also examine numerically the quality of the 
CIFT-reconstructed velocity field for such 
enhanced 
measurement configurations. 

The paper is structured as follows: after a short 
reminder of the basic principles and 
some numerical tests of CIFT for a typical 
LSC structure, we will
describe the experimental setup and 
present investigations of 
the dynamics of the azimuthal orientation of LSC 
for five specific temperature differences. 
Based on those measurements, we will discuss 
some preliminary 
results of the CIFT-reconstruction 
with our reduced sensor configuration. 

\section{Contactless inductive flow tomography and RB flows}
\label{cift}
In this section we discuss the mathematical basics
of CIFT and examine numerically how it can be
applied to typical RB convection problems.

\subsection{Basics of CIFT}

Here, we only briefly delineate 
the theory of CIFT. More details, 
in particular concerning 
the non-uniqueness problem
of the ill-posed inverse problem,
and how to mitigate it, can be found in 
previous papers 
\cite{Stefani2000,Stefani2000a}. 

Assume an electrically conductive fluid with 
a velocity field $\bf v$ exposed   
to a magnetic field $\bf B$.
According to Ohm's law in moving conductors, 
a current density
\begin{eqnarray}
  \label{eq:ohmsLawMovingCond}
  {\bf j} &=& \sigma({\bf v}\times {\bf B} - \nabla \varphi)
\end{eqnarray}
will be induced, with
$\sigma$ denoting the electrical
conductivity of the fluid and
$\varphi$ 
the electric scalar potential. 
Applying Biot-Savart's law \cite{Stefani2000}, 
this current density ${\bf j}$ induces the 
following  
(secondary) magnetic field ${\bf b}({\bf r})$:
\begin{eqnarray}
\label{eq:inducedBField}
    {\bf b}({\bf r})  &= &\frac{\mu_0\sigma}{4\pi}
    \iiint\limits_{V}  [{\bf v}({\bf r}') \times {\bf B}({\bf r}')]\times   
    \frac{{\bf r}-{\bf r}'}
  {|{\bf r}-{\bf r}'|^3}dV' \nonumber\\
  &&-\frac{\mu_0\sigma}{4\pi}\oiint\limits_{S} \varphi({\bf s}')
  {\bf n}({\bf s}')\times \frac{
    ({\bf r}-{\bf s}')} {|{\bf r}-{\bf s}'|^3}dS' \; .
\end{eqnarray}
It is important to   
incorporate the surface integral on the r.h.s. of 
Eq. (\ref{eq:inducedBField}) which, in certain circumstances, 
can completely cancel the volume integral term 
(early attempts \cite{Baumgartl1993,Berkov1995}   
to develop a CIFT-like magnetic flow tomography 
were flawed by this omission). 

Exploiting the divergence-free condition of $\bf j$,   
Eq. (\ref{eq:ohmsLawMovingCond}) leads to  
a Poisson equation for the
electric potential:
\begin{eqnarray}
\Delta \varphi&=&\nabla \cdot  ({\bf v}\times {\bf B}) \; .
\end{eqnarray} 
Using Green's theorem, the solution of 
this Poisson equation can be shown to fulfill the 
boundary integral equation
\begin{eqnarray}
  \label{eq:potentialEQ}
  \varphi({\bf s}) & = &\frac{1}{2\pi}\iiint\limits_{V}
  [{\bf v}({\bf r}') \times {\bf B}({\bf r}')]\cdot
  \frac{{\bf s}-{\bf r}'}
  {|{\bf s}-{\bf r}'|^3}dV'\nonumber\\
  &&-\frac{1}{2\pi}\oiint\limits_{S}\varphi({\bf s}'){\bf n}({\bf s}')\cdot\frac{
    ({\bf s}-{\bf s}')} {|{\bf s}-{\bf s}'|^3}dS'  \; ,
\end{eqnarray}
if insulating boundaries are assumed.

In general, the total magnetic field $\bf B$ under 
the integrals of Eqs. (\ref{eq:inducedBField}) and 
(\ref{eq:potentialEQ}) is the sum of an 
externally applied (primary) magnetic field ${\bf B}_0$ 
and the induced 
(secondary) magnetic field  $\bf b$ itself.
The ratio between $\bf b$ and ${\bf B}_0$ is 
proportional to the 
magnetic Reynolds number $Rm$ of the flow, defined as 
\begin{eqnarray}
  \label{eq:rm}
  Rm=\mu_0 \sigma l v \; ,
\end{eqnarray}
with $l$ and $v$ denoting characteristic length 
and velocity scales of the
fluid, respectively. For large values of $Rm$, and
appropriate flow topologies, it is possible to 
achieve self-excitation of a magnetic field.
In this case, one can obtain solutions of
Eqs. (\ref{eq:inducedBField}) and (\ref{eq:potentialEQ})
even for ${\bf B}_0=0$ , an effect that is known as 
homogeneous dynamo action \cite{StefaniRaedler2000,Xu2004}.

However, in most industrial applications, and in particular 
in typical RB convection problems, $Rm$ is 
typically smaller than $1$.
As for the experiment to be discussed here, 
with a vessel width of approximately 0.1 m, 
a typical LSC velocity of 
$v_{LSC}\approx 1$\,cm/s, and a 
conductivity of the liquid 
GaInSn of $3.27 \times 10^6$ S/m, we obtain  
$Rm \approx 0.004$. In such cases, it is well justified
to replace ${\bf B}$ by ${\bf B}_0$ under 
the integrals in Eqs.
(\ref{eq:inducedBField}) and (\ref{eq:potentialEQ}). 
Thereby, we arrive at a linear inverse 
problem for the determination of the velocity 
field $\bf v$ from the induced                                               
magnetic field $\bf b$ that is supposed to be 
measured in the exterior of the fluid. 
Note that the primary magnetic field can be
applied in various directions. As long as the 
typical time for velocity changes is
larger than the switching times for the 
different applied fields, this allows to collect
more information about (basically) the same 
velocity field which can further mitigate the
non-uniqueness of the inverse problem.
Examples of its solution in cylindrical geometry, and 
details on how to apply the Tikhonov regularization, can 
be found in \cite{Stefani2004,Wondrak2009}.

\subsection{Numerical tests of CIFT for RB flows}

In the following, we will discuss how CIFT can be
employed to infer the typical LSC flow structure 
of RB convection. 
For this purpose we have first simulated the RB flow 
using the {\it buoyantBoussinesqPimpleFoam} solver of the 
finite volume library OpenFOAM.
No turbulence models were used.
The time-average of the resulting flow served then
as the basis for computing 
the induced magnetic fields
according to Eqs. (\ref{eq:inducedBField}) and 
(\ref{eq:potentialEQ}). Finally, these fields 
were used, in turn, as input for reconstructing the 
velocity structure by solving the linear 
inverse problem. In this step, we tested
various  (hypothetical) configurations 
of magnetic field sensors situated around the 
vessel, and applied also different combinations 
of primary magnetic fields.

\begin{figure}[!htbp]
	\centering
	\includegraphics[width=0.8\textwidth]{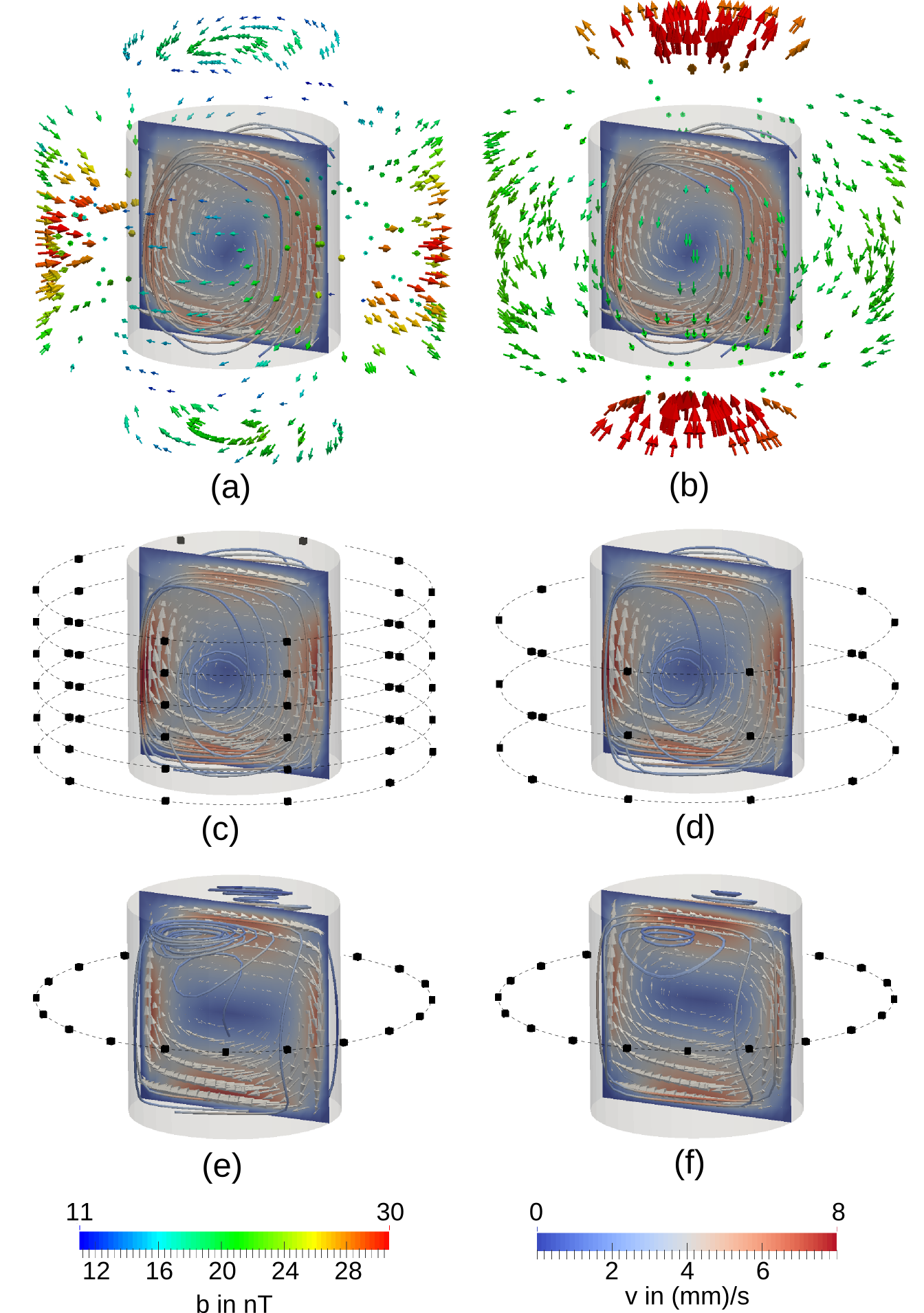}
	\caption{
	Numerical tests of CIFT for an RB flow at $\Delta T=2.2$\,K.
	(a) OpenFOAM-simulated velocity structure, and the induced
	magnetic field for an applied vertical field. (b) As in (a), but 
	for an applied horizontal field. (c) Measurement configuration with
	6$\times$10 sensors, and CIFT inferred velocity for 
	applying both a vertical and a horizontal field. (d) As in 
	(c), but for 3$\times$10 sensors. 
	(e) As in (c), but for 1$\times$20 sensors. (f) As in (e), 
	but restricted to 
	measurements  with only applied vertical field. 
	}
	\label{Fig:numerics}
\end{figure}

Figs.~\ref{Fig:numerics}a,b show the OpenFOAM-simulated, time-averaged 
velocity field for a temperature difference 
of $2.3$\,K, together with the flow-induced magnetic field 
for applied vertical field (a) and applied horizontal field (b),
and the CIFT-reconstructed velocity field by
assuming four different measuring configurations 
(c,d,e,f). For the inversion shown in Fig.~\ref{Fig:messung2grad}c we 
assume homogeneous  
vertical and horizontal primary fields of $3$\,mT to be 
applied, and a hypothetical number of 6$\times$10 sensors
distributed over 6 heights and 10 azimuths 
around the mantle of the cylinder.
The visual impression of an excellent coincidence
between the original (Fig.~\ref{Fig:numerics}a) 
and the reconstructed flow (Fig.~\ref{Fig:numerics}c) 
is supported by an empirical correlation coefficient
as high as  0.95 and a relative rms error  as low as 
0.09. A similar coincidence 
holds if the number
of sensors is reduced to $3\times 10$ as 
in Fig.~\ref{Fig:numerics}d.
Then the inversion ends up  with an 
empirical correlation coefficient
of 0.95 and a relative rms error  of 0.11.

Amazingly, even the strongly  reduced sensor
configuration as used in our experiment, with
20 sensors at mid-height, distributed regularly 
along the azimuth, provides reasonable results.
When applying both a vertical and a 
horizontal field (Fig.~\ref{Fig:numerics}e) we still 
obtain an empirical correlation coefficient
of 0.84 and a relative rms error  of 0.29.
The minor deterioration of the agreement 
seems to be connected with
the appearance of an (artificial) eddy structure at the 
top, where 
some necessary magnetic field information is obviously 
missing.
Even when restricting the measurements to the case
of a vertically applied field (Fig.~\ref{Fig:numerics}f), 
we still get a reasonable empirical correlation coefficient 
of 0.8 and a relative rms error of 0.36.
This surprisingly robust behavior of CIFT suggests 
to start our RB investigation with the 
reduced measurement configuration as
shown in Fig.~\ref{Fig:numerics}f.

\section{Experimental setup}
\label{setup}

The setup of our RB experiment (Fig.~\ref{Fig:setup})
consists of a cylindrical 
column filled with the
eutectic liquid alloy GaInSn that is 
homogeneously heated from below. 
In trying to emulate the growing crystal of a Czochralski 
system, only partial cooling of the top is realized 
by means of a circular heat exchanger mounted 
concentrically within the upper lid. This partial 
cooling covers approximately the same relative area 
as the growing crystal in a real Cz-puller. Both heat 
exchangers (heater and cooler) are made of copper 
with branched channels to minimize any remaining 
temperature gradients. PID-controlled thermostats 
with a large reservoir were used to supply both 
the top heat exchanger and the bottom plate with 
the coolant/heating fluid at high flow rate. Fig.~\ref{Fig:setup}b 
shows a photograph of the 
experimental cell with the diameter and height of 
$D=H=87$\,mm (aspect ratio $\Gamma=D/H=1$). The 
transparent side wall is made of borosilicate 
glass, because of its low heat conductivity. During the
measurements, the whole apparatus is
embedded in mineral wool in order to minimize lateral 
heat loss (Fig.~\ref{Fig:setup}c).

For a given temperature difference $\Delta T$ between
bottom and top, we obtain an (approximate) 
Rayleigh number of
\begin{eqnarray}
Ra&=&\frac{g \beta}{\kappa \nu} H^3 \Delta T \nonumber\\
&=&\frac{g \beta \rho c_p}{\lambda \nu} H^3 \Delta T \nonumber\\
&=&3.42 \times 10^8 \frac{H^3 \Delta T}{m^3 K} \nonumber\\
&=&2.25 \times 10^5 \frac{\Delta T}{K} 
\end{eqnarray}
by considering $g=9.81$\,m/s$^2$ and the following
material parameters of GaInSn at $20^{\circ}$\,C \cite{Plevachuk2015}:
thermal expansion coefficient $\beta=1.24 \times 10^{-4}$\,K$^{-1}$,
density $\rho=6350$\,kg/m$^3$,
specific heat capacity at constant pressure 
$c_p=366$\,J/(kg k),
viscosity $\nu=3.44 \times 10^{-7}$\,m$^2$/s, thermal
diffusivity $\lambda=24.0$\,W/(m K). Note that, in 
particular at high $\Delta T$, $Ra$ should be 
determined more accurately by taking into account the
temperature dependence of material parameters 
(which here are simply 
fixed to a reference temperature of 20$^{\circ}$\,C).
Note also that, even for the same value of $Ra$, 
one should not expect 
a perfect agreement with other liquid metal 
RB experiments, 
because of the partial cooling at the top. 

In the first experimental realization of CIFT, 
the excitation magnetic field
is generated  by only one pair of circular coils 
which are fed by a current of 20 A generating in 
the fluid a fairly homogeneous vertical field $B_z$ 
of approximately 4 mT. 
In asking whether this field itself could have an 
influence on the flow structure, 
we determine the Hartmann number as
$Ha=B_z H (\sigma/(\rho \nu)^{1/2} \approx 13$,
and the maximum expectable 
interaction parameter (estimated
for the lowest LSC velocity of $v_{LSC} \approx 0.5 $\,cm/s) 
as $N=Ha^2/Re \approx 0.14$. Both values suggest
that the influence of the measuring field on the 
flow should be minor. 

The coils are cooled in order to minimize temperature 
changes over time. In contrast to the more advanced 
measurement configurations as 
numerically examined in the previous section, 
the flow induced magnetic field 
is measured by only 20 Fluxgate probes which are 
regularly distributed over the azimuth, approximately 
at mid-height of the RB vessel. The measured 
magnetic field component 
is always the radial one, pointing normally to the
nearest surface point of the vessel.
For low $\Delta T$, the measured 
induced magnetic field is typically in the order
of $10...100$\,nT, which is a factor 
10$^{-5}$ smaller than 
the primary magnetic field! For this reason
it is essential to guarantee a very stiff 
mounting of the sensors with respect to the coils 
in order to minimize bias magnetic field 
perturbations that would result
from any deformations due to 
thermal expansion of the coil and/or the sensor 
holders.

Complementary to the magnetic field measurements, 
the temperature was measured 
at the top of the vessel, slightly offset from the
center (see Fig.~\ref{Fig:setup}a).

\section{Experiments and flow reconstructions}
\label{experimentandreco}

In this section we present some exemplary 
magnetic field measurements, and discuss 
the possibility to employ CIFT on the basis of 
our reduced sensor configuration.

After carefully taking into account 
various sources of measurement errors, 
in particular the thermal expansion of the setup, and 
redesigning the measurement setup accordingly, 
we were able to measure the very weak 
LSC induced magnetic fields for temperature 
differences reaching from 2.2\,K to 80.8\,K 
(corresponding to Rayleigh numbers 
between $5.18 \times 10^5$ and $1.82 \times 10^7$).

Most of the experiments were carried out during night, 
in order to avoid magnetic field perturbations which 
inevitably result from 
any moving ferromagnetic parts or from switching 
high-amperage currents during typical
daytime operations in our experimental 
hall (see \cite{Ratajczak2015} for illustrations).

\subsection{Experiments at $\Delta T=2.2$\,K}
\label{2komma2}

\begin{figure}[!htbp]
	\centering
	\includegraphics[width=0.6\textwidth]{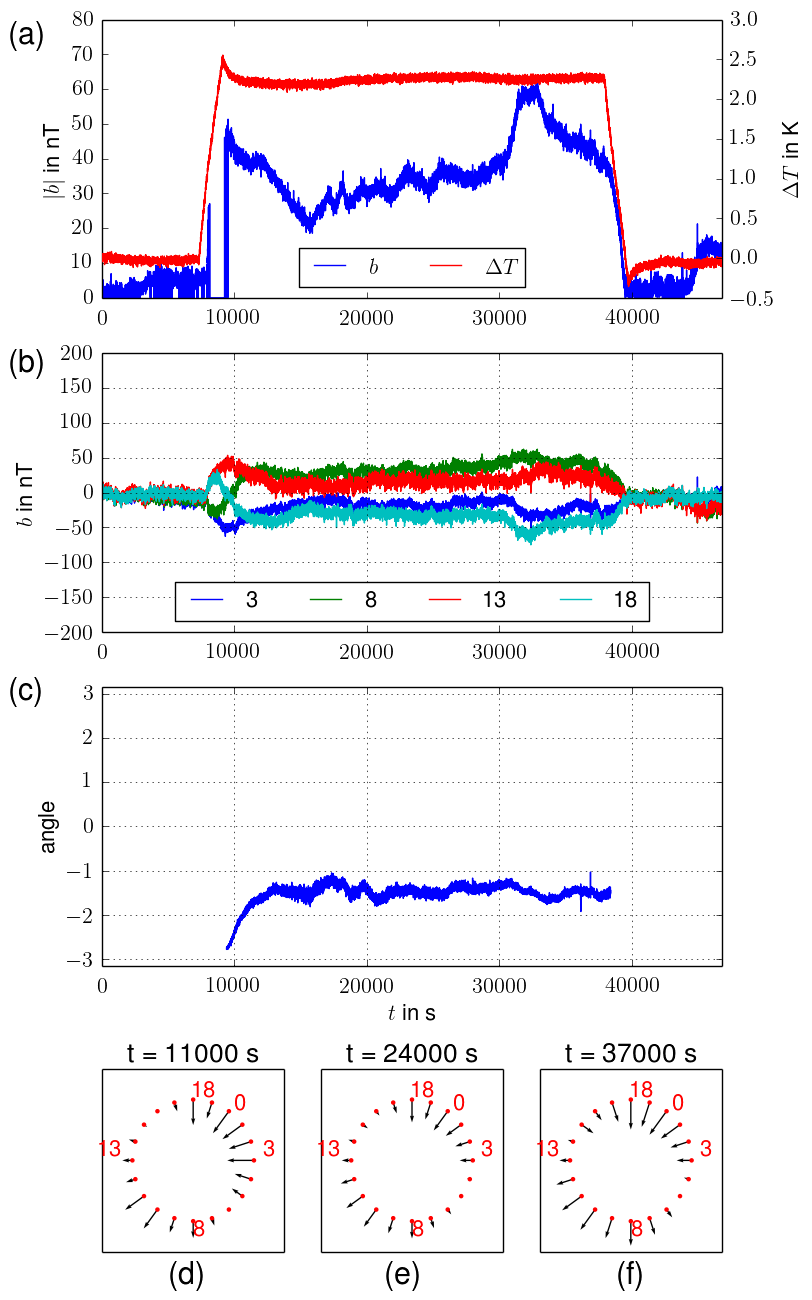}
	\caption{
	Measurements for $\Delta T=2.2$\,K. (a) Temperature difference
	and amplitude $|b(t)|$ of the sine fit 
	to the 20 measured magnetic fields. 
	(b) Signals of sensors 3,8,13,18
	which are separated by 90$^{\circ}$. 
	(c) Inferred angle 
	$\phi_0(t)$. (d,e,f) All 20 magnetic 
	signals as measured at three
	different snapshots.}
	\label{Fig:messung2grad}
\end{figure}

Fig.~\ref{Fig:messung2grad} illustrates a typical 
measurement of the flow induced 
magnetic fields, and some derived features, 
for an average temperature difference $\Delta T=2.2$\,K. 
First, Fig.~\ref{Fig:messung2grad}a shows the 
actual temperature difference together with
the amplitude $|b|$ of a sine function 
$|b(t)| \sin(\phi -\phi_0(t))$ 
fitted to the 20 magnetic field
data in dependence on time. Here, $\phi$ denotes the 
angle as measured in clockwise direction starting from
sensor No. 0 (see Fig.~\ref{Fig:messung2grad}d,e,f). After establishing the 
temperature difference at $t \approx 10000$\,s, the flow induced 
magnetic field rises and is maintained until the 
temperature difference is set to zero again at $t\approx 40000$\,s.
Second, Fig.~\ref{Fig:messung2grad}b shows the signal 
of 4 selected sensors which are positioned in quadrature. 
The sine fit gives also the time-dependent offset angle
$\phi_0(t)$, which lays in the middle between the
angles of dominant upward and downward 
flow of the LSC 
(Fig.~\ref{Fig:messung2grad}c). The angle where the 
LSC can be expected to have its maximum upward  
velocity, can be determined as $\phi_{up}=\phi_0 + \pi/4$.
Finally, Figs.~\ref{Fig:messung2grad}d,e show the structure 
of the measured 
magnetic field at three selected 
instants $t=11000$\,s, $t=24000$\,s, and $t=37000$\,s.

For this comparably low temperature difference, 
we obtain relatively smooth data 
which indicate a fairly stable LSC that is
basically pinned 
at a certain angle, around which it performs
weak oscillations. Such a pinning of the LSC 
at some particular angle had also been 
observed by Cioni et al.~\cite{Cioni1997}
who had attributed it to weak inhomogeneities
of the experimental set-up (in particular, the
heating system).

\begin{figure}[!htbp]
	\centering
	\includegraphics[width=0.6\textwidth]{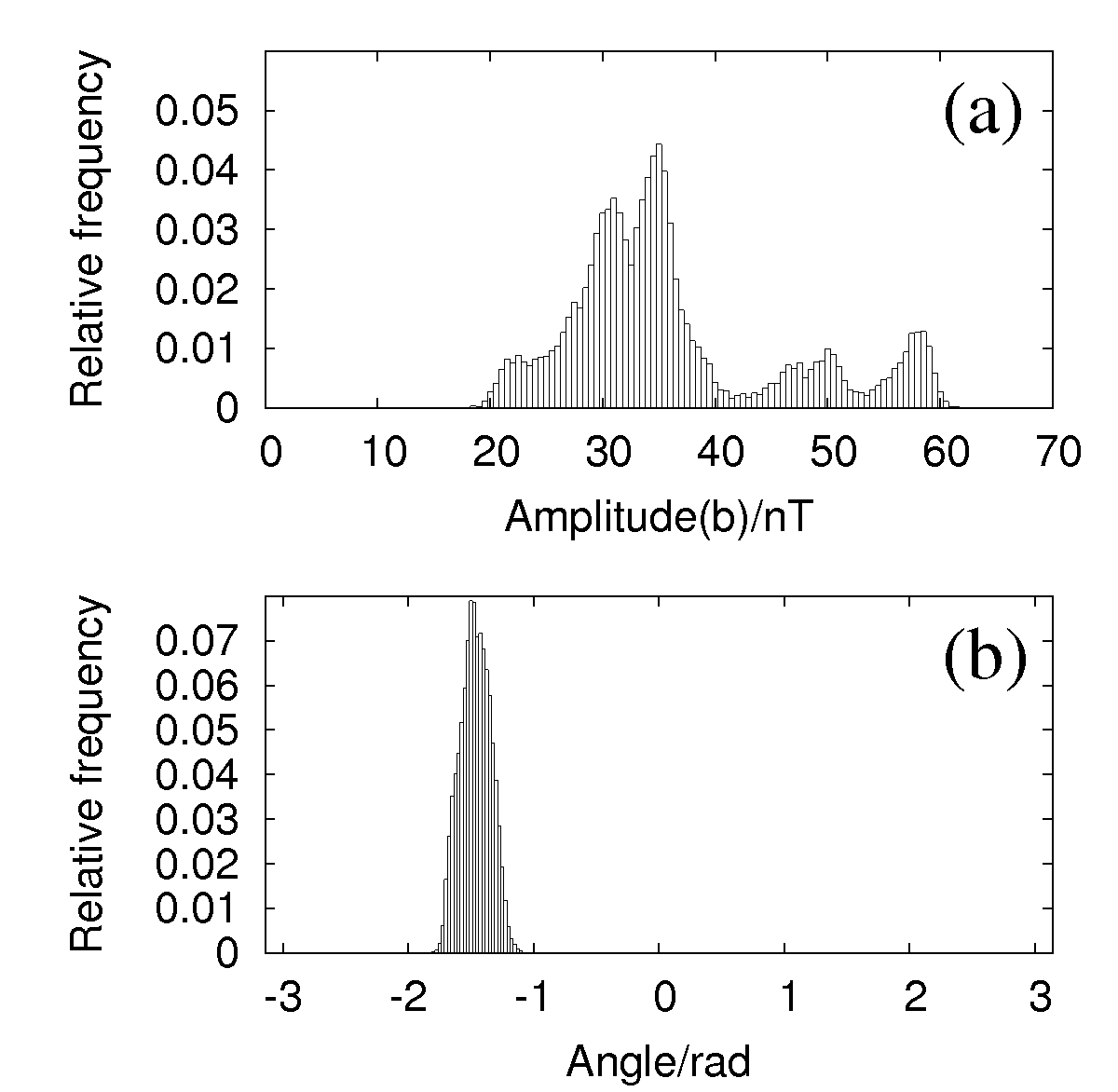}
	\caption{
	Histogram of the amplitude of the sine fit 
	to the 20  measured magnetic fields (a), and
	of the angle $\phi_0$ of the LSC (b), 
	for $\Delta T=2.2$\,K.}
	\label{Fig:histo2grad}
\end{figure}

The impression of a ''quiet'' LSC is also supported by 
Fig.~\ref{Fig:histo2grad} 
which shows the histograms of the amplitude $|b(t)|$
of the 20 
measured magnetic fields (a), and the histogram of 
the angle $\phi_0(t)$ (b).

\begin{figure}[!htbp]
	\centering
	\includegraphics[width=0.6\textwidth]{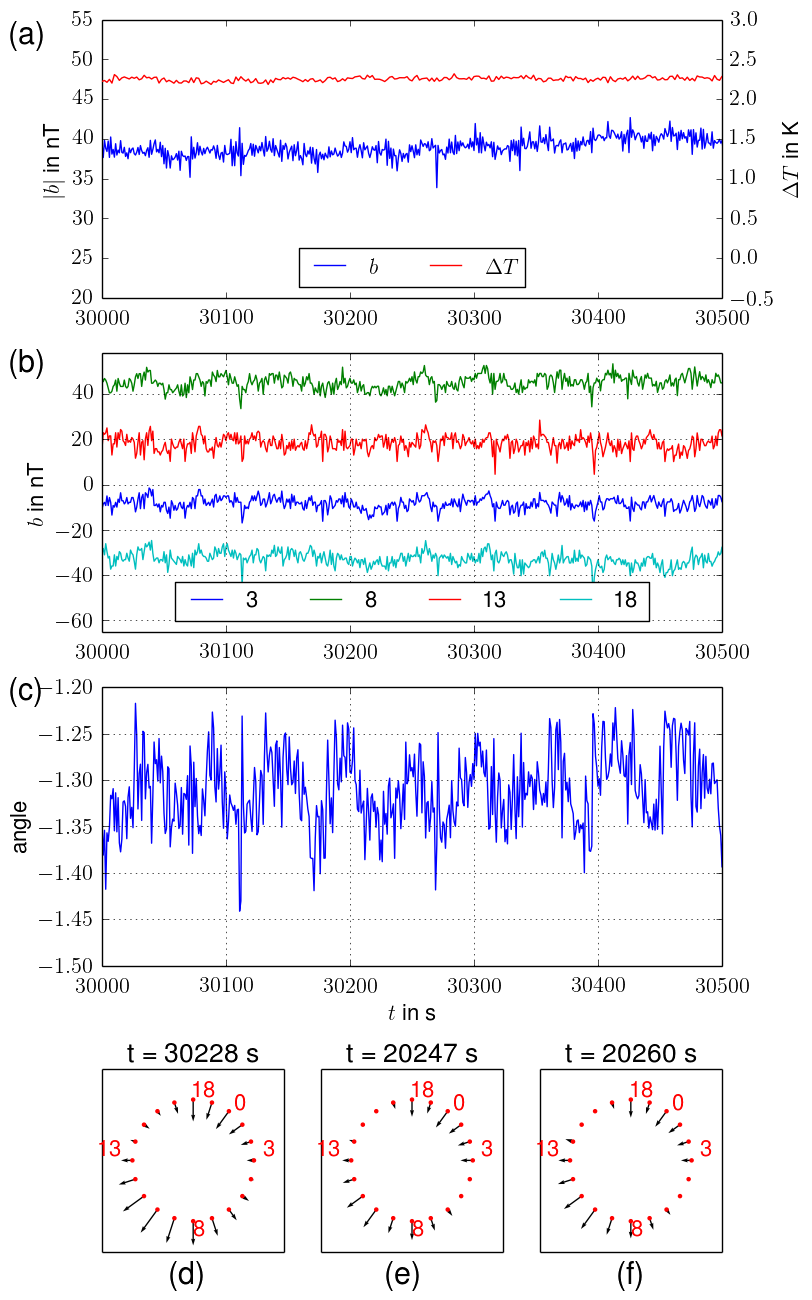}
	\caption{
	As in Fig.~\ref{Fig:messung2grad}, but restricted to the
	period $30000...30500$\,s. A dominant oscillation with a frequency
	of 0.0185 Hz is clearly visible in the sensor and angle data.}
	\label{Fig:messung2graddetails}
\end{figure}

Fig.~\ref{Fig:messung2graddetails} is an enlarged section 
of Fig.~\ref{Fig:messung2grad} for the
time-interval $30000...30500$\,s. 
Sensors 8 and 13, in particular, show a regular 
oscillation, whose sharp frequency $f=0.0185$\,Hz 
can be identified in the Fourier transform 
of the time series (see Fig.~\ref{Fig:fourier} 
further below). This frequency, which lays in the 
typical range of the turn-over frequency 
$f_{turnover} \approx v_{LSC}/(4 H)$ 
of the LSC, very likely corresponds 
to either a torsional
or a sloshing mode \cite{BrownAhlers2009}.
Given the restriction of our sensor data 
to only one height, we refrain here from trying
to distinguish between the two modes,
and leave that for future work.

\subsection{Experiments at $\Delta T=5.3$\,K}
\label{5komma3}

\begin{figure}[!htbp]
	\centering
	\includegraphics[width=0.6\textwidth]{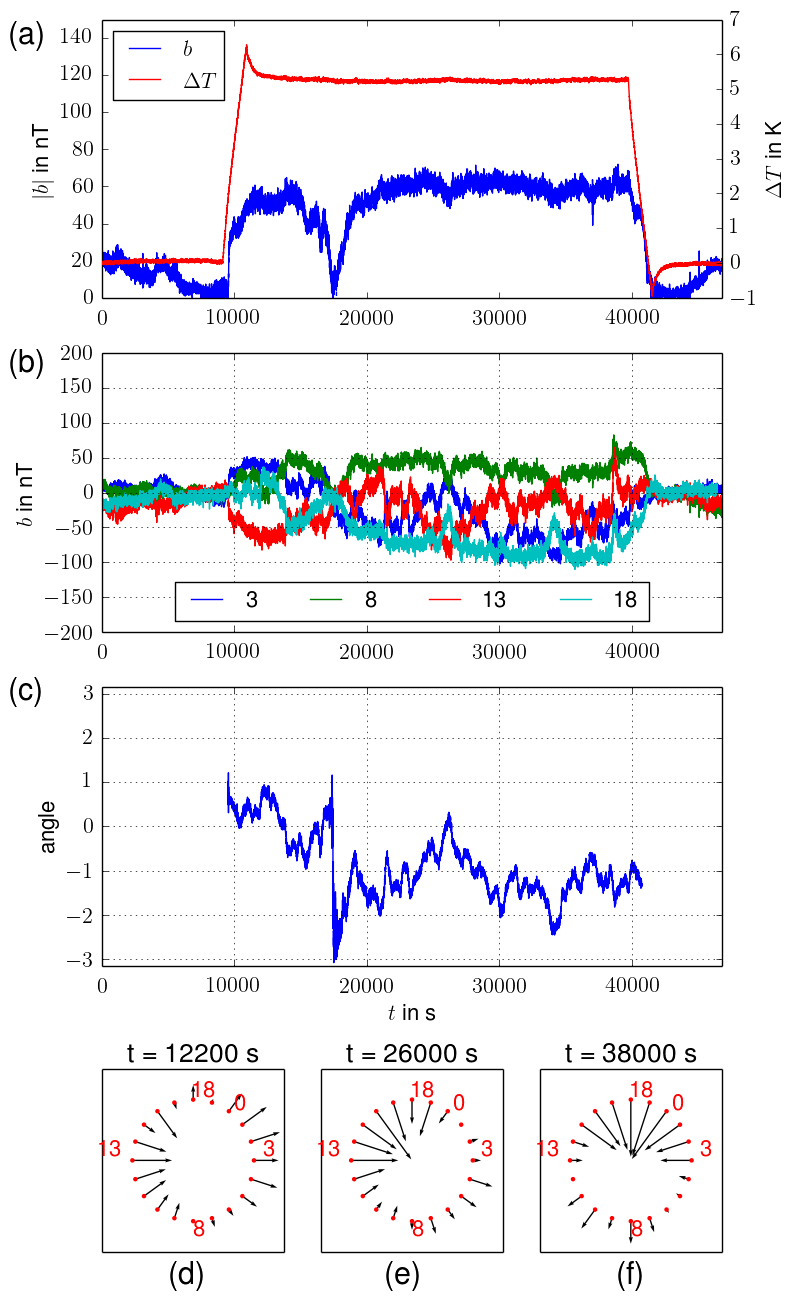}
	\caption{
	As in Fig.~\ref{Fig:messung2grad}, but for 
	$\Delta T=5.3$\,K.}
	\label{Fig:messung5grad}
\end{figure}

\begin{figure}[!htbp]
	\centering
	\includegraphics[width=0.6\textwidth]{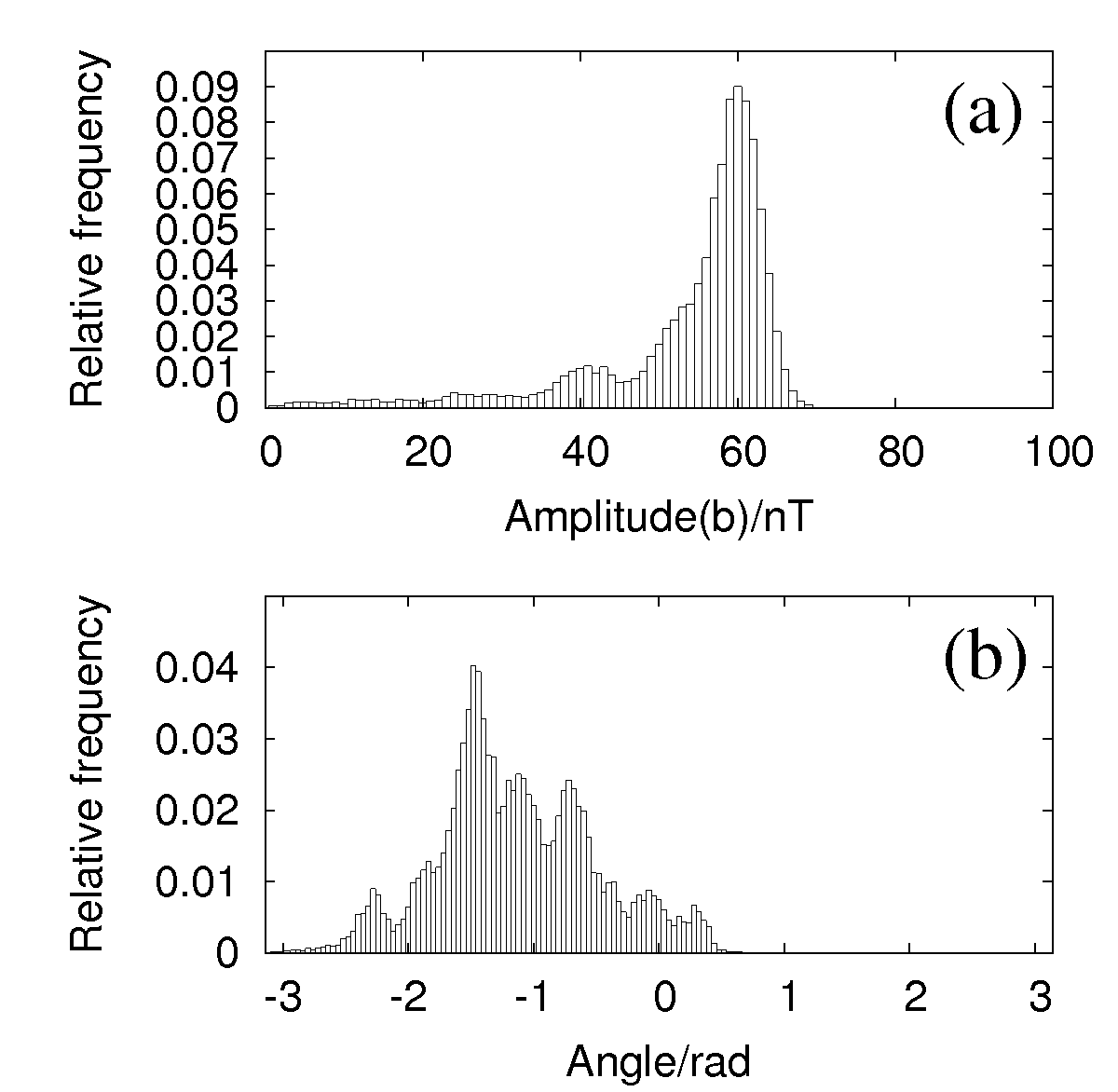}
	\caption{As in Fig.~\ref{Fig:histo2grad}, but for 
	$\Delta T=5.3$\,K.}
	\label{Fig:histo5grad}
\end{figure}

New effects come into play when increasing the 
temperature difference to $\Delta T=5.3$\,K.
Fig.~\ref{Fig:messung5grad} shows again the complete 
time-line of the experiment, which 
reveals significantly stronger variations of the 
LSC. This is also visible in the histograms 
(Fig.~\ref{Fig:histo5grad}), in particular in the broader
distribution of the amplitude.

\begin{figure}[!htbp]
	\centering
	\includegraphics[width=0.6\textwidth]{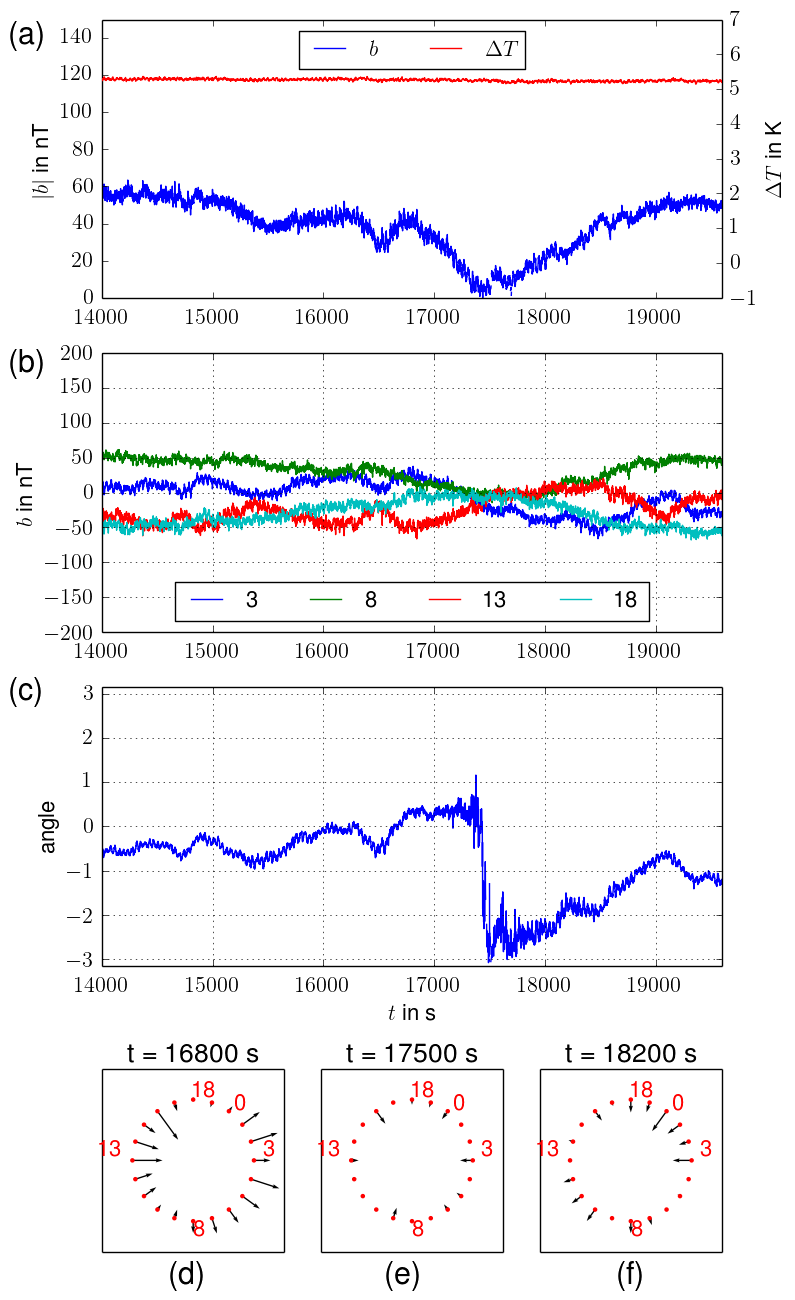}
	\caption{
	As in Fig.~\ref{Fig:messung5grad}, but restricted to the
	period $14000...19600$\,s.  The flow ceases at around $t=17500$\,s.}
	\label{Fig:messung5graddetails}
\end{figure}

An interesting detail of this experiment is exhibited 
in Fig.~\ref{Fig:messung5graddetails} which covers the 
time interval $14000...19600$\,s. Around $t=17500$s we observe
a vanishing of the amplitude of the magnetic fields,
which clearly indicates a ''cessation'' 
of the LSC as it had been described previously 
by various authors \cite{BrownAhlers2006,Xi2008}.
After this cessation, the LSC starts again at another 
angle (Fig.~\ref{Fig:messung5graddetails}c).

\subsection{Experiments at $\Delta T=12.5$\,K}
\label{12komma5}

\begin{figure}[!htbp]
	\centering
	\includegraphics[width=0.6\textwidth]{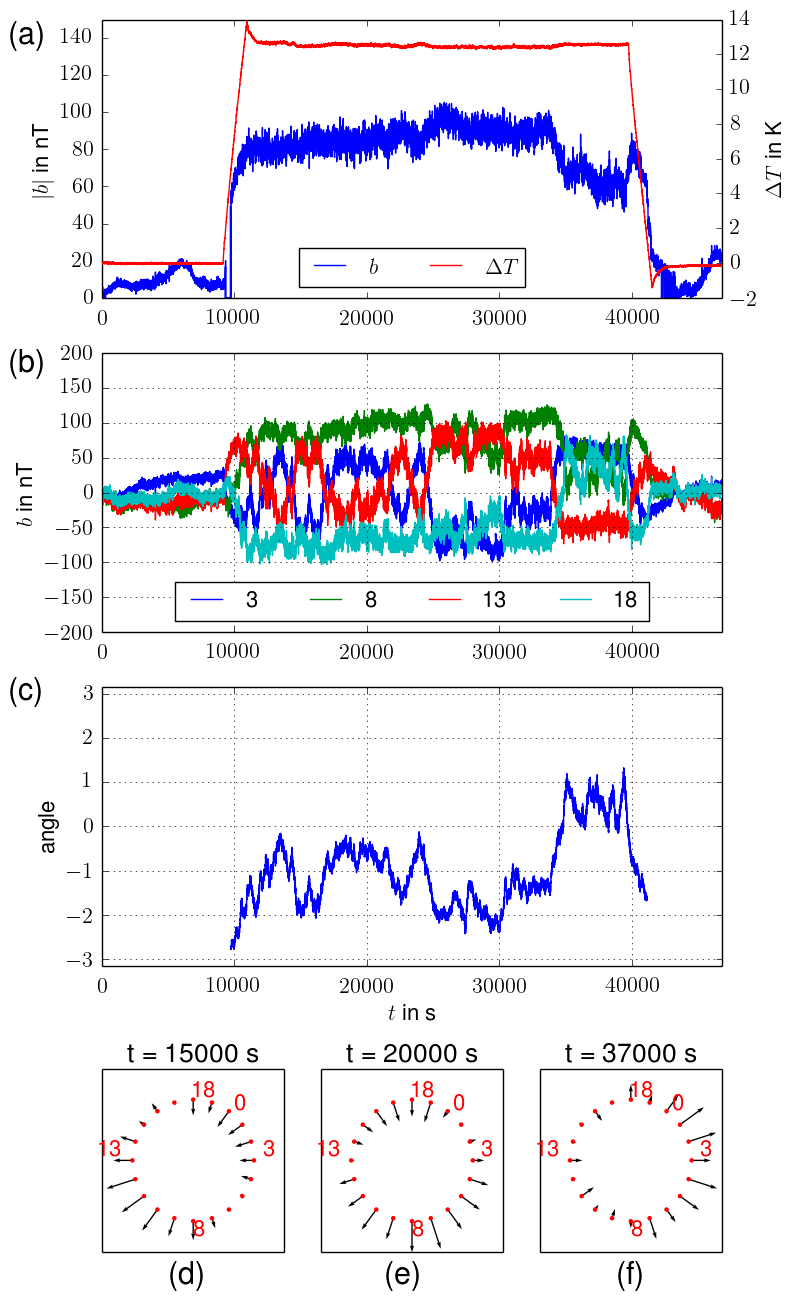}
	\caption{
	As in Fig.~\ref{Fig:messung2grad}, but for 
	$\Delta T=12.5$\,K.}
	\label{Fig:messung12grad}
\end{figure}

A somewhat different behavior is observed at 
$\Delta T=12.5$\,K.
Fig.~\ref{Fig:messung12grad}a indicates a rather 
constant amplitude of the fields (and hence of the 
intensity of
the LSC), with just a moderate reduction close to 
the end of the experiment.
However, the individual sensor signals (Fig.~\ref{Fig:messung12grad}b) 
point to strong fluctuations of the angle of the LSC, 
which is quantified in Fig \ref{Fig:messung12grad}c and illustrated 
in the three snapshots of Figs.~\ref{Fig:messung12grad}d,e.

\begin{figure}[!htbp]
 	\centering
	\includegraphics[width=0.6\textwidth]{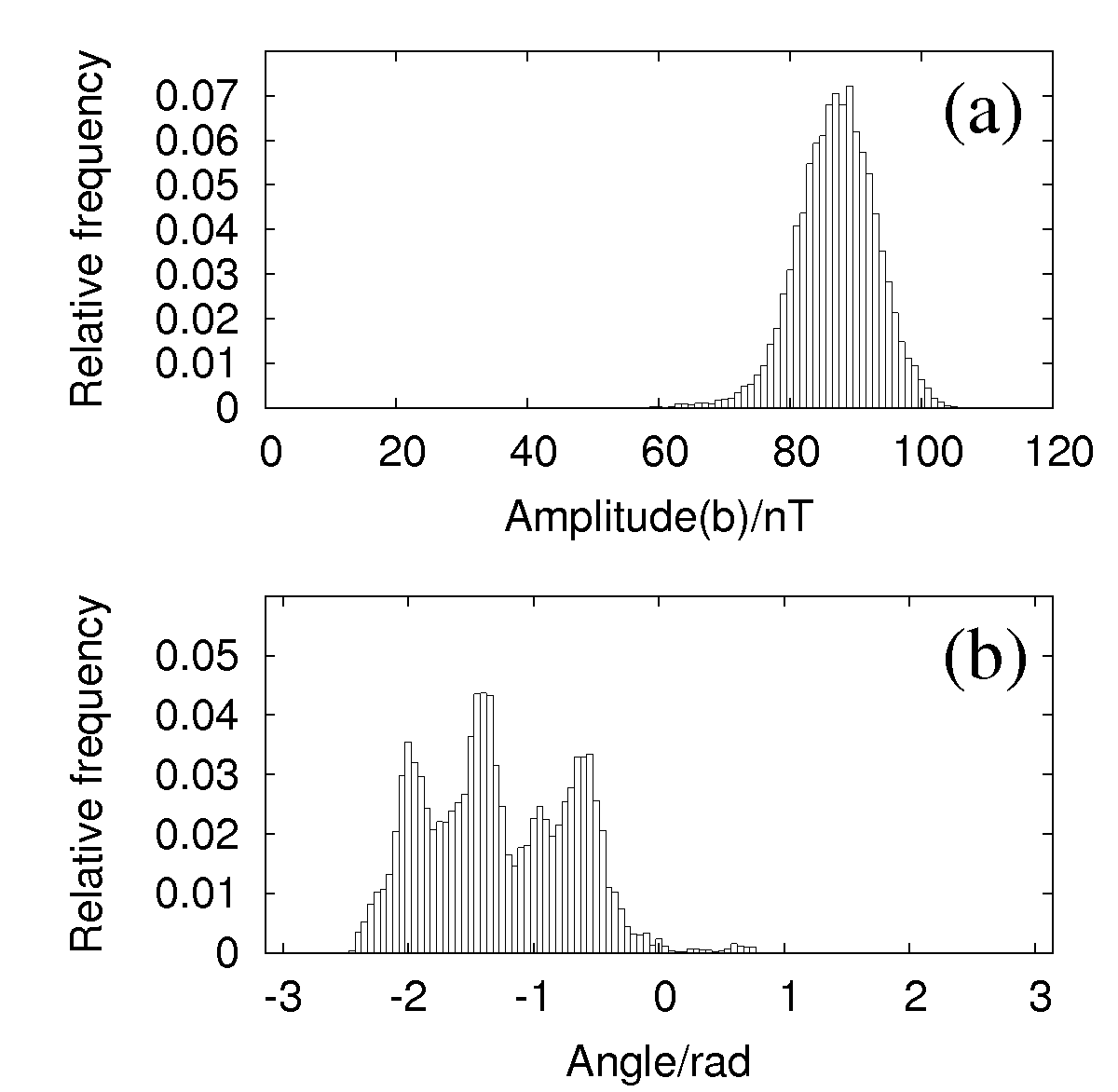}
	\caption{As in Fig.~\ref{Fig:histo2grad}, but for 
	$\Delta T=12.5$\,K. }
	\label{Fig:histo12grad}
\end{figure}

The corresponding histograms show indeed a 
remarkable constant amplitude (Fig.~\ref{Fig:histo12grad}a) 
and a kind of tri-modal distribution of the angle 
(Fig.~\ref{Fig:histo12grad}b).

\begin{figure}[!htbp]
	\centering
	\includegraphics[width=0.6\textwidth]{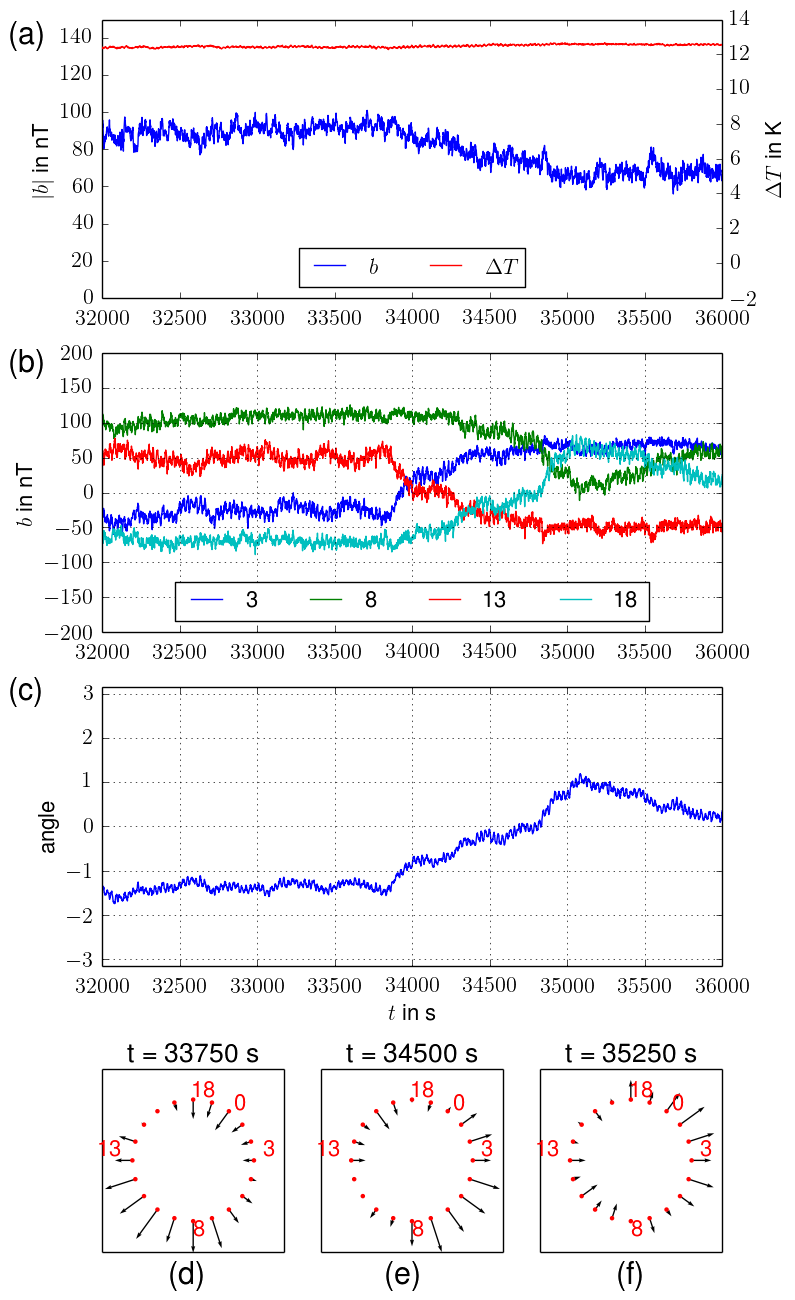}
	\caption{
	As in Fig.~\ref{Fig:messung12grad}, but restricted to the
	period $32000...36000$\,s. Note the rotation of the angle at around 
	$t=34500$\,s. }
	\label{Fig:messung12graddetails}
\end{figure}

A rather slow rotation of the angle by around 180$^{\circ}$
can be identified in the details for the time-interval 
$32000...36000$\,s, see Fig.~\ref{Fig:messung12graddetails}.

\subsection{Experiments at $\Delta T=31.6$\,K}
\label{31komma6}

\begin{figure}[!htbp]
	\centering
	\includegraphics[width=0.6\textwidth]{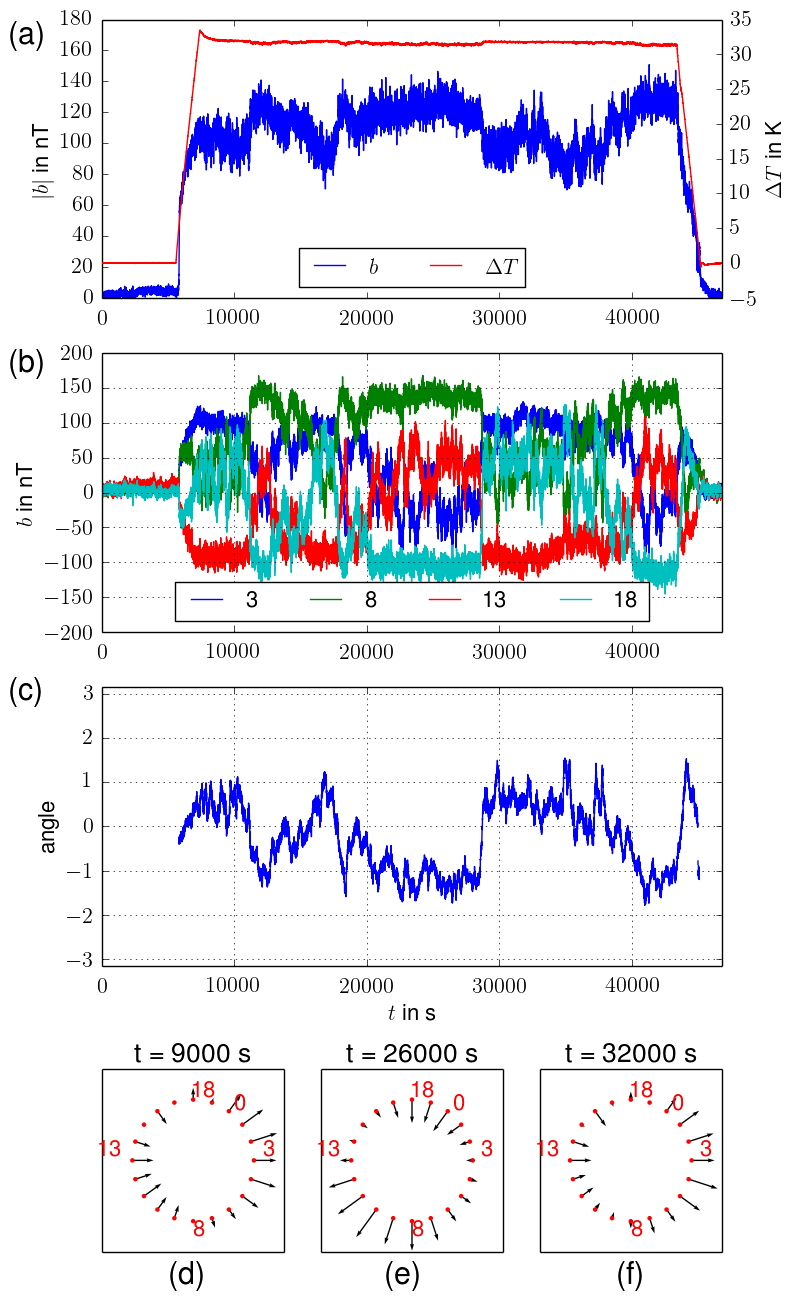}
	\caption{
	As in Fig.~\ref{Fig:messung2grad}, but for 
	$\Delta T=31.6$\,K.}
	\label{Fig:messung32grad}
\end{figure}

\begin{figure}[!htbp]
	\centering
	\includegraphics[width=0.6\textwidth]{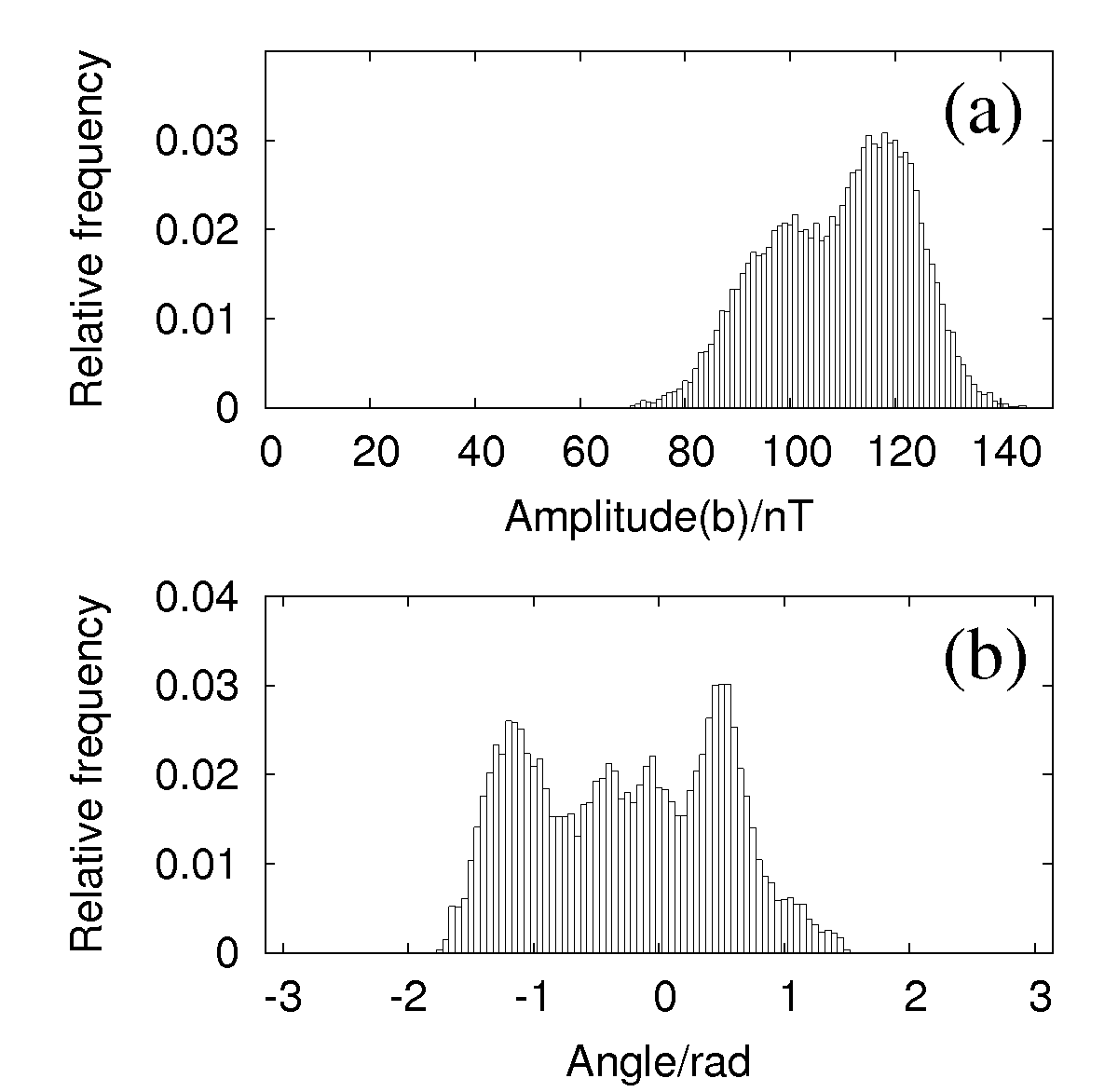}
	\caption{As in Fig.~\ref{Fig:histo2grad}, but for 
	$\Delta T=31.6$\,K.}
	\label{Fig:histo32grad}
\end{figure}

Yet another feature appears at $\Delta T=31.6$\,K
(Fig.~\ref{Fig:messung32grad}).
Again, the amplitude of the field remains rather constant, while
the individual fields and the angle undergo sudden
jumps and reversals. The histogram (Fig.~\ref{Fig:histo32grad}b) 
shows now a sort
of bi-modal behavior of the angle with two
dominant angles which are opposite to each other.

\begin{figure}[!htbp]
	\centering
	\includegraphics[width=0.6\textwidth]{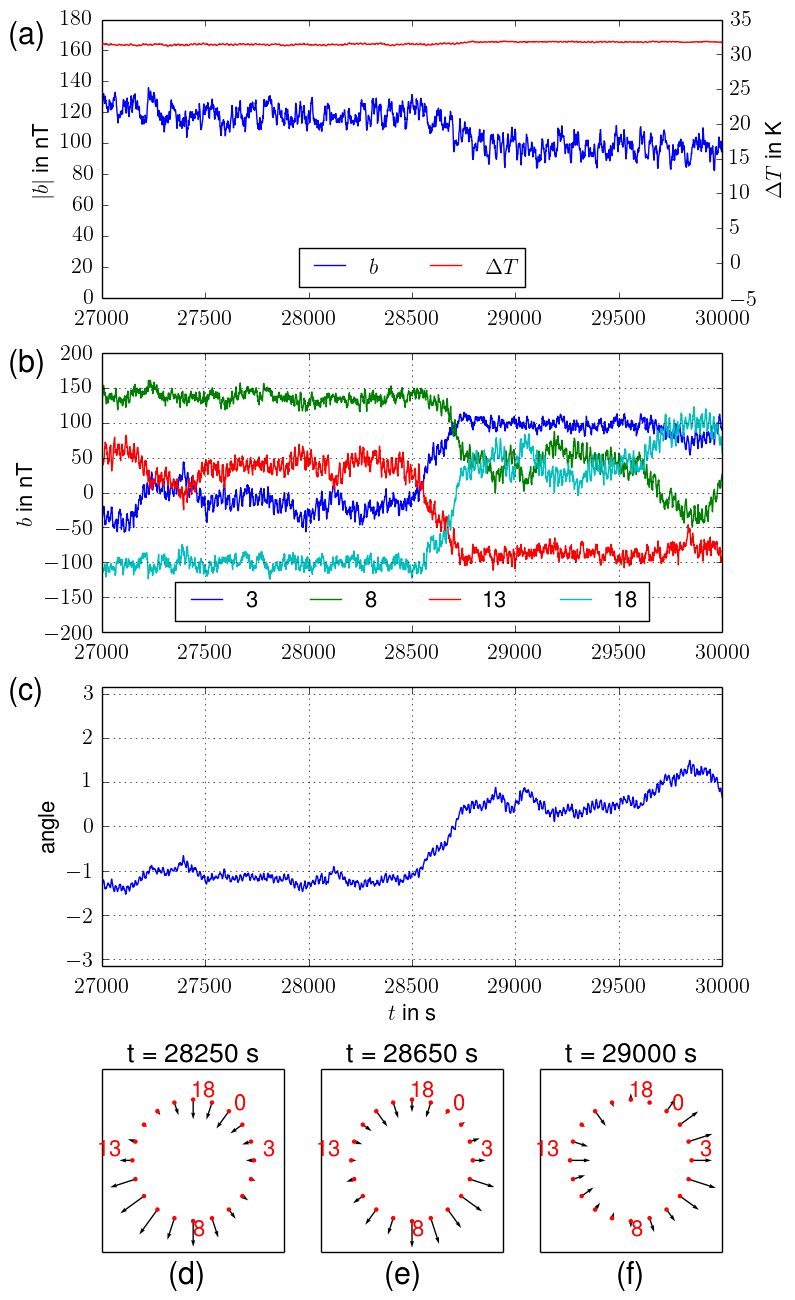}
	\caption{
	As in Fig.~\ref{Fig:messung32grad}, but restricted to the
	period $27000...30000$\,s. Note the fast 
	rotation of the angle starting at 
	$t=28500$\,s. }
	\label{Fig:messung32graddetails}
\end{figure}

A typical reversal between the two angles, which 
relies on a fast rotation instead of a cessation, 
is illustrated in detail in 
Fig.~\ref{Fig:messung32graddetails}.

\subsection{Experiments at $\Delta T=80.8$\,K}
\label{80komma8}

\begin{figure}[!htbp]
	\centering
	\includegraphics[width=0.6\textwidth]{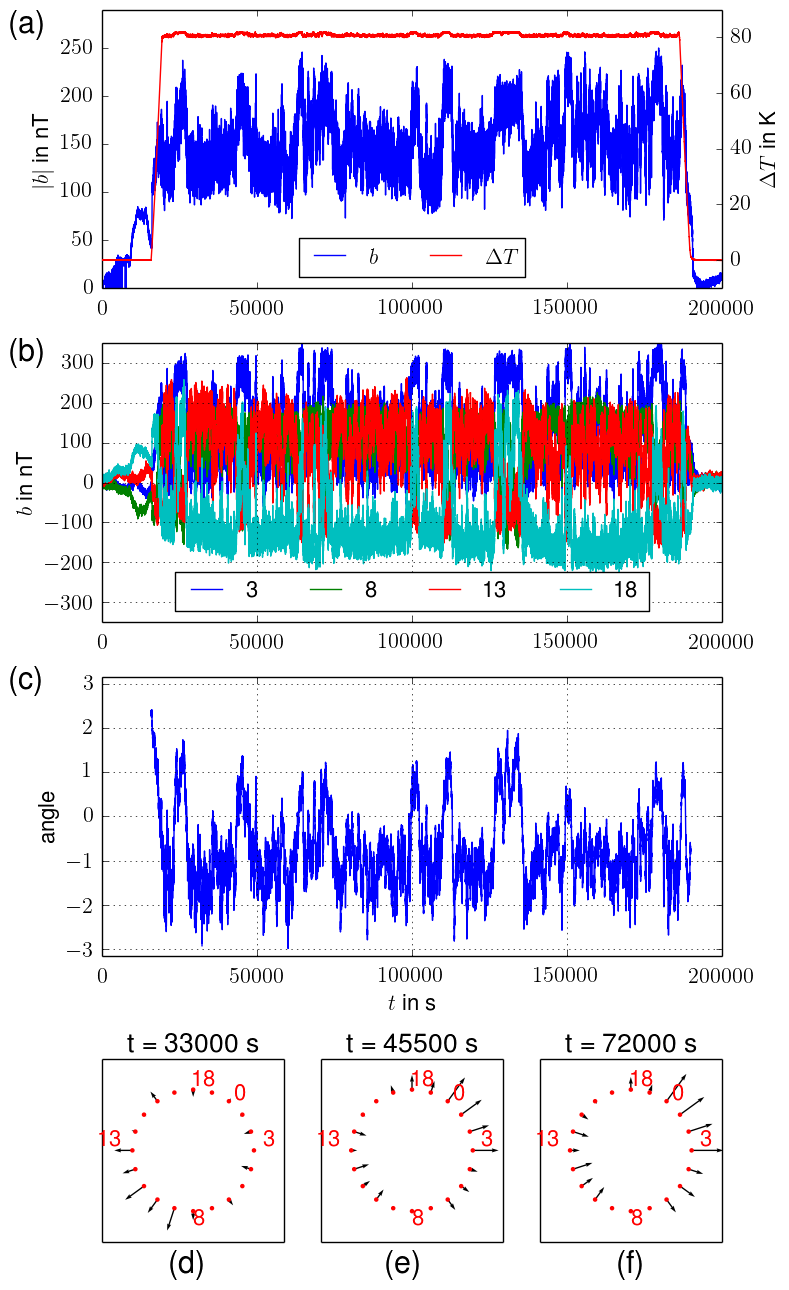}
	\caption{
	As in Fig.~\ref{Fig:messung2grad}, but for 
	$\Delta T=80.8$\,K.}
	\label{Fig:messung80grad}
\end{figure}

\begin{figure}[!htbp]
	\centering
	\includegraphics[width=0.6\textwidth]{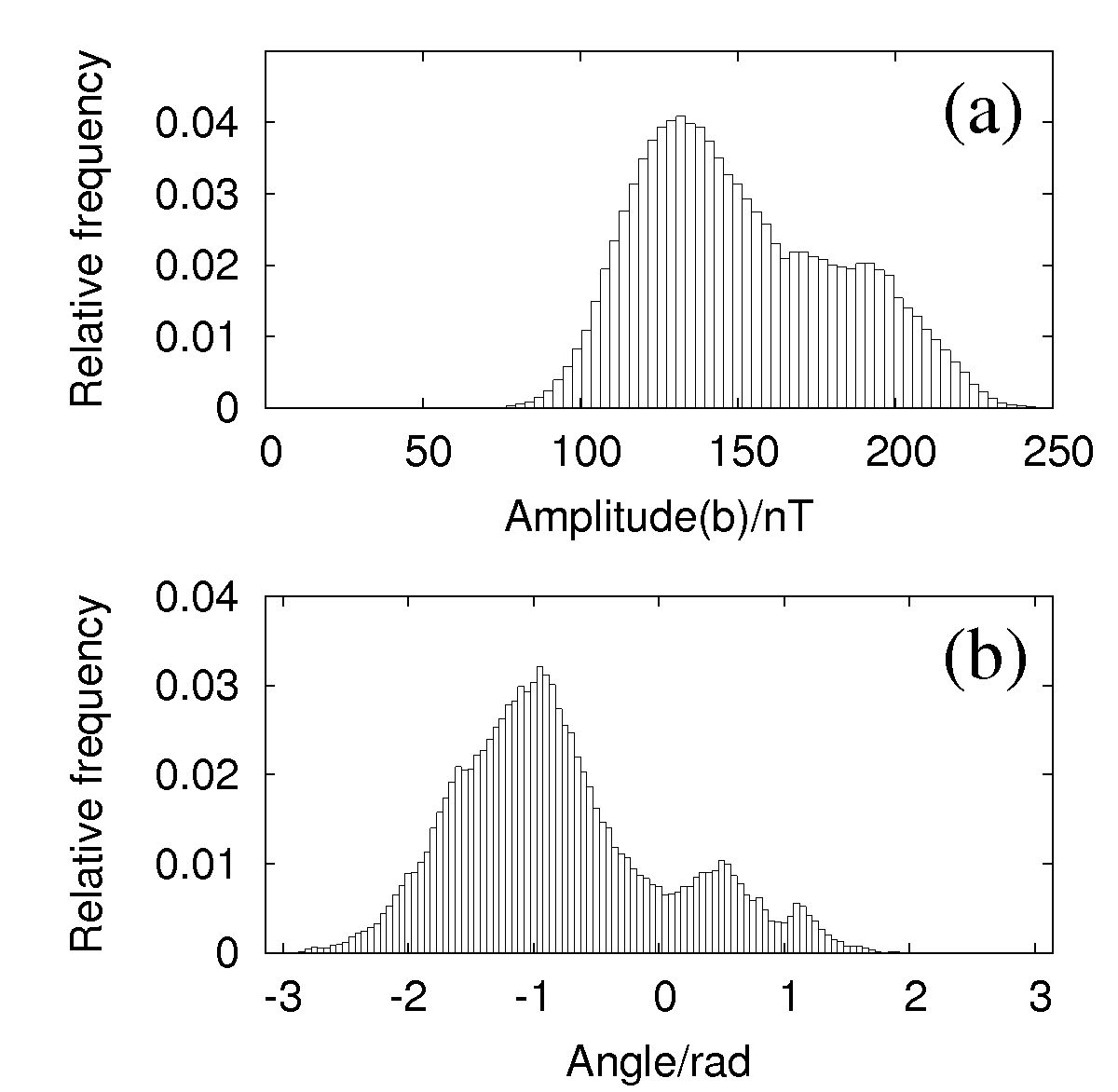}
	\caption{As in Fig.~\ref{Fig:histo2grad}, but for 
	$\Delta T=80.8$\,K.}
	\label{Fig:histo80grad}
\end{figure}

In contrast to the shorter experiments discussed
above, the run at $\Delta T=80.8$\,K has been carried 
out over a time interval as long as $50$\,h.
Over this period, the amplitude of the fields 
does not change very much, while
the individual fields and the angle of the LSC undergo 
again strong variations (Fig.~\ref{Fig:messung80grad}). 
The histogram 
(Fig.~\ref{Fig:histo80grad}) shows now a sort
of bi-modal behavior of both the amplitude and
the angle, which on closer inspection even seem to be
correlated.

\begin{figure}[!htbp]
	\centering
	\includegraphics[width=0.6\textwidth]{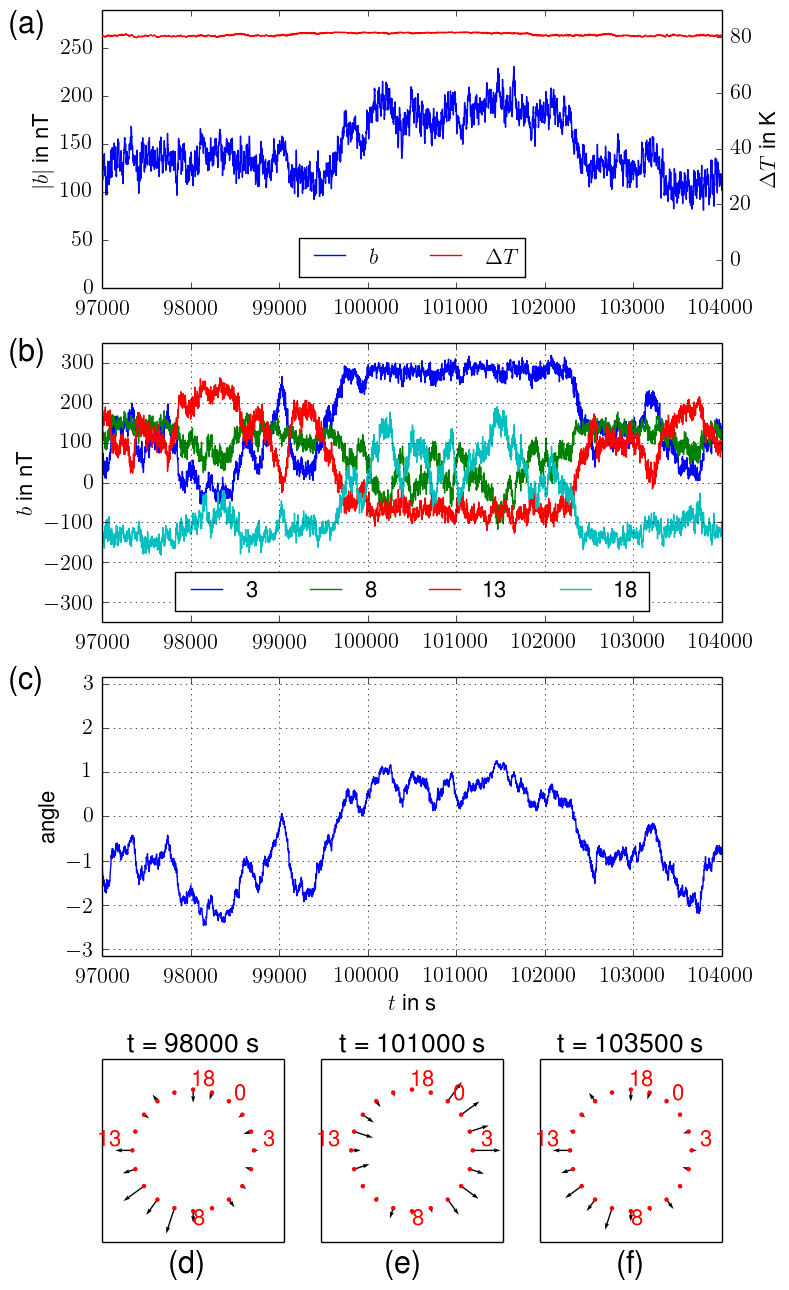}
	\caption{
	As in Fig.~\ref{Fig:messung80grad}, but restricted to the
	period $97000...10400$\,s.  Note the ''excursion'' of the 
	angle during this period.}
	\label{Fig:messung80graddetails}
\end{figure}

Fig.~\ref{Fig:messung80graddetails} shows the details of 
an ''excursion'' of the angle of the LSC.

\subsubsection{Summary of experiments}
\label{summary}

\begin{figure}[!htbp]
	\centering
	\includegraphics[width=0.95\textwidth]{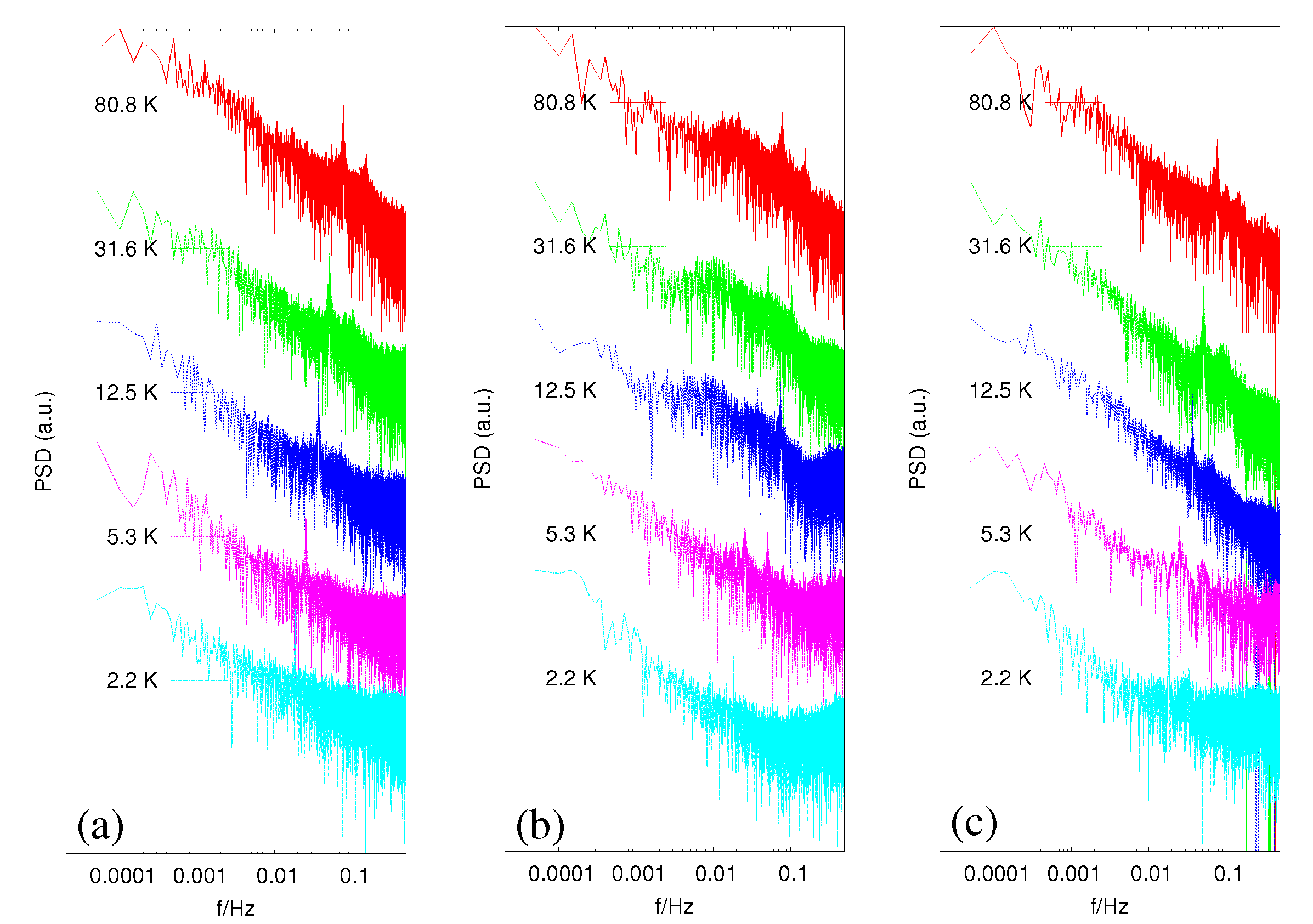}
	\caption{Spectra for the five runs with 
	different temperature differences.
	(a) Spectra of the signal of sensor 8. (b) Spectra 
	of the amplitude of the sine fit to the 
	20  measured signals. (c) Spectra 
	of the angle of the LSC as determined from
	the 20 sensors. Note that the signals underlying all
	spectra were equally restricted to the interval
	$t=15000...35000$\,s (even if the long measurements 
	at $\Delta T=80.8$\,K would 
	have allowed for a much longer interval). The PSD is
	given logarithmically in arbitrary units, and
	the individual curves have been shifted 
	appropriately for the sake of better visibility.}
	\label{Fig:fourier}
\end{figure}

For five different values of $\Delta T$ 
(and, therefore, of $Ra$) the measured magnetic 
field perturbations have unveiled a variety of 
flow phenomena.
These include a LSC that is pinned at some
angle from were it undergoes slight oscillations 
which, very likely, correspond to
torsional and/or sloshing modes. 
In all experiments 
the frequencies of these modes turned out 
to be quite sharp. This is illustrated
in Fig.~\ref{Fig:fourier} which shows the
spectra for the signal of sensor
8 (a), for the amplitude of all 20 
sensor data (b), and for the 
angle $\phi_0$ of the LSC (c). Whereas the main peaks 
of the torsional/sloshing mode are clearly visible in 
all three features, some
interesting distinctions can be made.
For example, the peak of the amplitude for $\Delta T=12.5$\,K
is rather weak, while the corresponding peak of the
angle is both strong and sharp. This is in agreement with
the histograms in Fig.~\ref{Fig:histo12grad}. Somewhat 
opposite to this is the behavior at $\Delta T=5.3$\,K,
which shows a weaker oscillation of the angle, but a 
more pronounced 
fluctuation of the amplitude 
(see also Fig.~\ref{Fig:histo5grad}).

\begin{figure}[!htbp]
	\centering
	\includegraphics[width=0.6\textwidth]{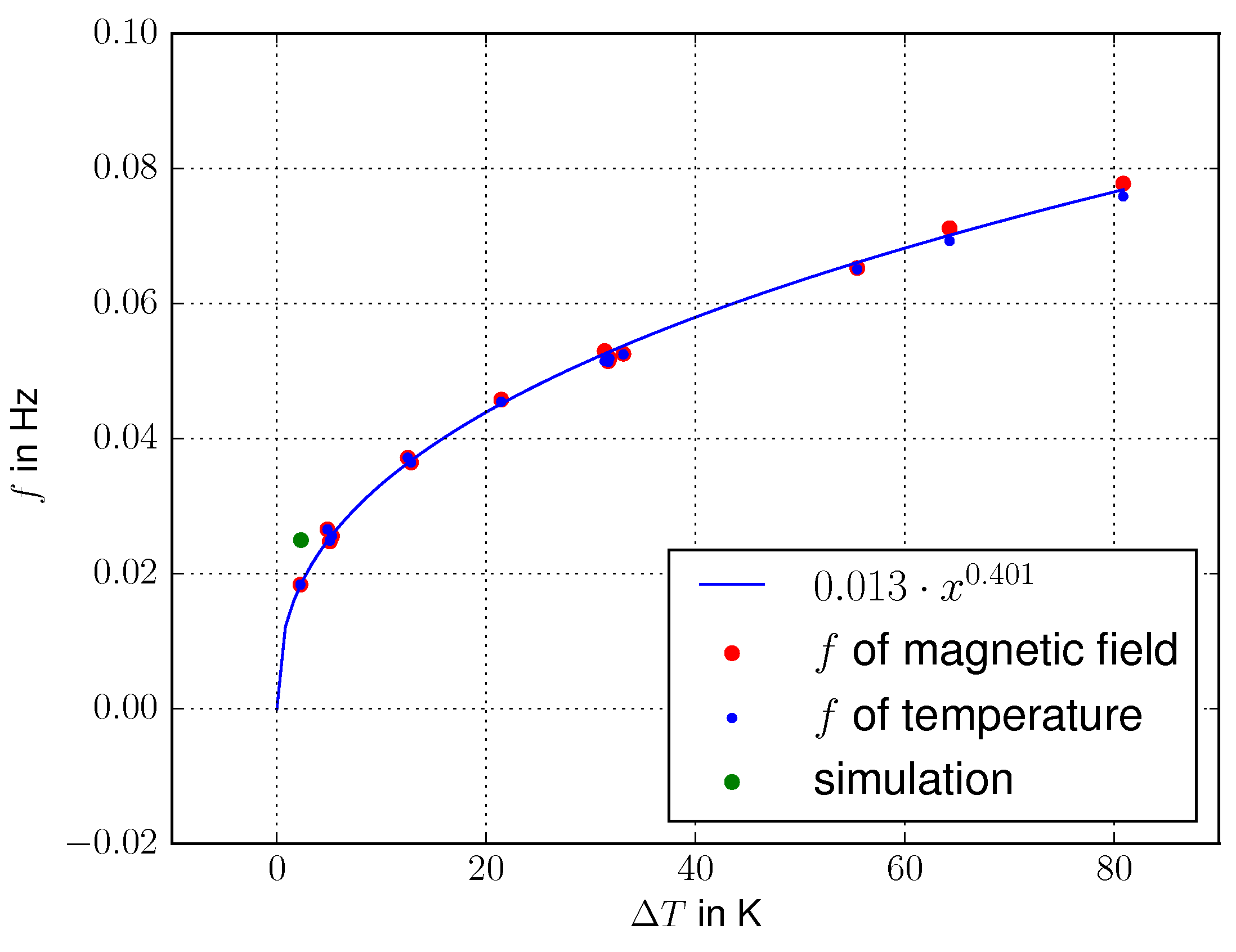}
	\caption{Dependence of measured frequencies 
	on $\Delta T$, and the corresponding fit curve 
	$f=0.013 \Delta T^{0.401}$.
	The measurements were done simultaneously with 
	the CIFT system (red) and with temperature sensors
	(blue). One single simulated
	frequency (green) fits well into the overall picture.
	}
	\label{Fig:frequency}
\end{figure}

Based on a total of 14 experiments (from which we have discussed
only five in detail), the dependence of the dominant 
frequency on $\Delta T$ is summarized in 
Fig.~\ref{Fig:frequency}. 
The fit curve of these values 
is $f=0.013 \Delta T^{0.401}$ which 
turns out to be in reasonable agreement with a
corresponding result 
of Cioni et al. \cite{Cioni1997} who had derived a relation
$2\pi f=0.47 \kappa Ra^{0.424}$. The remaining 
difference of the exponent might be due to the 
partial cooling of the top, and/or to
the temperature dependence
of various material parameters which modifies, 
in particular for
higher $\Delta T$, the
simplified linear 
relation between $\Delta T$ and $Ra$.

Apart from this general dominance of the high frequency 
modes, 
the experiments at larger $\Delta T$ show further 
effects, such as one (single !) cessation as observed 
for 
$\Delta T=5.3$\,K, and the more frequent reversals
at even higher $\Delta T$. The additional 
''bulge'' of the amplitude spectrum 
around $f\sim 0.01$\,Hz, which becomes visible for 
$\Delta T=31.6$\,K and $\Delta T=80.8$\,K (Fig.~\ref{Fig:fourier}b),
indicates the appearance of low-frequency fluctuations 
of the LSC intensity, perhaps even a
bi-modal behavior as also suggested in
Fig.~\ref{Fig:histo80grad}a.

\subsection{Flow reconstructions}
\label{reco}

The presented measurements show that, despite of
their weakness, the flow 
induced magnetic fields can be safely measured, and
that the time-dependence of the LSC  
can be reliably determined. 
While a full three-dimensional reconstruction
of the flow would require a 
second primary field to be applied
in horizontal direction, 
and also more sensors to be 
distributed at different heights,
we can still try to reconstruct the
flow from just the 20 measurements at hand.

\begin{figure}[!htbp]
	\centering
	\includegraphics[width=0.28\textwidth]{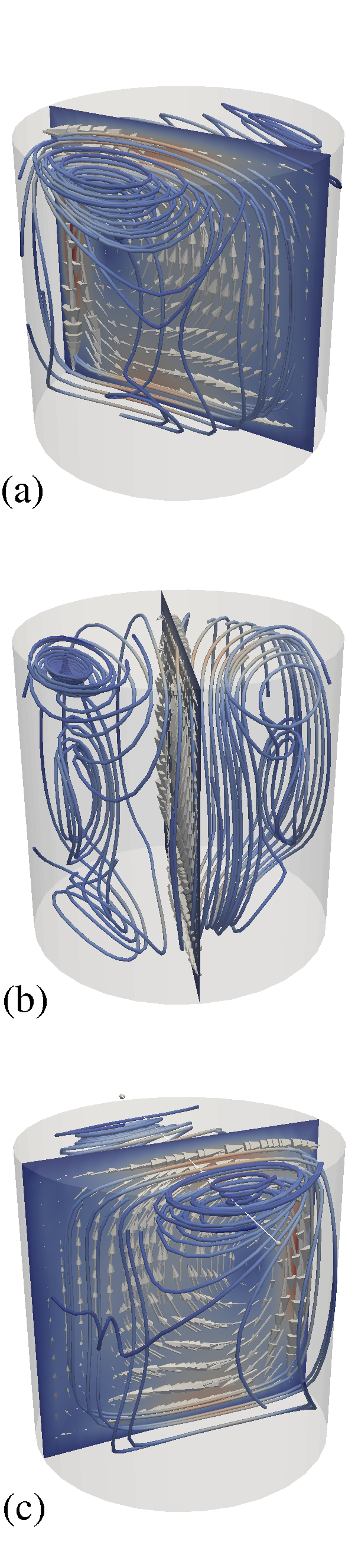}
	\caption{
	CIFT-reconstructed velocity for $\Delta T=31.6$\,K,
	at  the instants $t=28250$\,s (a), $t=28650$\,s (b),
	and $t=29000$\,s (b), corresponding to the three
	measurements shown in Figs.~\ref{Fig:messung32graddetails}d,e,f, 
        respectively.}
	\label{Fig:reko}
\end{figure}

As an example we use here the three data snapshots from
Figs.~\ref{Fig:messung32graddetails}d,e,f, whose 
CIFT-reconstructions are shown now in Figs.~\ref{Fig:reko}a,b,c, 
respectively. 
Apart from some unrealistic eddy structures at 
the top (which are likely artifacts due to the 
missing magnetic field information at that height as seen in 
Figs.~\ref{Fig:numerics}g,f),
the global rotation of the LSC becomes clearly visible.

\section{Conclusions}
\label{conclusions}

In this paper, we have demonstrated the 
viability of CIFT for measuring 
the LSC in liquid metal convection
without any contact with the melt 
and without disturbing the homogeneous 
thermal boundary conditions at top and bottom. 
Even by only measuring
the flow induced magnetic fields for one single
(vertical) applied magnetic,
it was possible to identify the direction and the
dynamics of the LSC. 

If a second magnetic field were applied to the 
convecting liquid, the full three-dimensional, 
time-dependent 
flow in the cylinder 
could be reconstructed with even better quality, 
although perhaps not with the very high 
empirical correlation coefficient as obtained, 
in the numerical part 
of this paper, for the time-averaged velocity. 
For this enhanced measurement configuration, 
we still have to 
find an optimum number and spatial distribution 
of sensors. 
One of the purposes of such 
an enhancement would also be to distinguish 
between torsional and sloshing modes of the 
LSC. The application of low-frequency AC fields, and 
the use of (gradiometric) pick-up coils,
could further improve the robustness and 
technical applicability of the method.

\section*{Acknowledgment}

Financial support of this research by the 
German Helmholtz Association in the frame of 
the Helmholtz-Alliance LIMTECH is gratefully 
acknowledged.





\end{document}